# NepoIP/MM: Towards Accurate Biomolecular Simulation with a Machine Learning/Molecular Mechanics Model Incorporating Polarization Effects

Ge Song

*Department of Chemistry, Duke University, Durham, NC 27708, USA*

Weitao Yang*

*Department of Chemistry and Department of Physics, Duke University, Durham, NC 27708, USA*
E-mail: weitao.yang@duke.edu

**Abstract**

Machine learning force fields offer the ability to simulate biomolecules with quantum mechanical accuracy while significantly reducing computational costs, attracting growing attention in biophysics. Meanwhile, leveraging the efficiency of molecular mechanics in modeling solvent molecules and long-range interactions, a hybrid machine learning/molecular mechanics (ML/MM) model offers a more realistic approach to describing complex biomolecular systems in solution. However, multiscale models with electrostatic embedding require accounting for the polarization of the ML region induced by the MM environment. To address this, we adapt the state-of-the-art NequIP architecture into a polarizable machine learning force field, NepoIP, enabling the modeling of polarization effects based on the external electrostatic potential. We found that the nanosecond MD simulations based on NepoIP/MM are stable for the periodic solvated dipeptide system and the converged sampling shows excellent agreement with the reference QM/MM level. Moreover, we show that a single NepoIP model can be transferable across different MM force fields, as well as extremely different MM environment of water and proteins, laying the foundation for developing a general machine learning biomolecular force field to be used in ML/MM with electrostatic embedding.

## 1. Introduction

To describe and simulate the dynamic behavior of biomolecules, empirical physics model known as Molecular Mechanics (MM) force fields have been actively developed and widely used for several decades. With the advancements in hardware today, computer simulation of biomolecules based on MM force fields is highly efficient and has successfully uncovered the mechanism of many biological processes with atomistic details (1-10). However, there is still a need for improving systematically the accuracy of simulations, as they often fail to capture the correct conformational features of diverse biomolecules, particularly those with flexible structures (11-14).

The accuracy of classical molecular dynamics is limited by the simple function form of its underlying MM force fields. Although hybrid Quantum Mechanics/Molecular Mechanics (QM/MM) approaches have been developed to enhance the modeling accuracy of the QM subsystem of interest (15, 16), the intensive calculations required by QM theory significantly restrict the timescale and lengthscale of QM/MM MD simulations. Typically, the QM subsystem is selected to be only tens of atoms, such as the active site of an enzyme (17), and the computation time for this region can already be many magnitudes longer than that of the MM surroundings. Therefore, even for small QM subsystems, it is prohibitively computational-intensive to reveal their dynamic behaviors by QM/MM MD simulations on a nanosecond timescale.

The extensive development of neural network architectures for molecular potentials in recent years has provided an opportunity to address this computational dilemma, as neural networks can, in principle, be fitted to arbitrary QM theories and make much faster predictions (18-26). However, directly applying them to biomolecular systems is still problematic (27). First of all, biomolecules surrounded by solvent and ions form a large and complex system, making it difficult to build a training dataset that adequately covers the whole system's physical space with QM calculations. Even if such a dataset is constructed to train the neural network, modeling the whole system with neural networks is hard to be as computationally efficient and stable as the MM force fields (27, 28), since the latter has an extremely clear and simple form to approximate the long range interaction. Therefore, a Machine Learning/Molecular Mechanics (ML/MM) multiscale model, parallel to QM/MM, is of great interest in the context, which raises the issue of the development of an embedding scheme.

The simplest embedding scheme is the mechanical embedding, i.e. the internal energy of ML region remains the same as it being in vacuo and the interaction between ML and MM atoms is treated at the MM level using standard Coulomb and Lennard-Jones potentials (29). Studies such as the NNP/MM model (30) and the machine learning protein force field, Charmm-NN (31), are examples of the mechanical embedding. While is simple, its major limitation is that the polarization, or the change in the energy from an isolated molecule to the same molecule within MM environment, is entirely ignored. Such polarization effects are important even for classical MM force field development: there is already a long history of using the 'pre-polarized' point charges fitted against the 6-31G* electrostatic potential to implicitly address such an inconsistency caused by the water environment (32).

In the commonly used QM/MM electrostatic embedding scheme, where the electrostatic interactions between the QM electron density and MM charges are included in the QM hamiltonian, the polarization effects have been explicitly accounted for (33, 34). To implement such an embedding scheme in ML/MM, two routes have emerged.

The first route is to input descriptors of the MM environment to the neural network to learn the polarization effects. A straightforward strategy is to allow the models to incorporate atom-wise features from MM atoms within the cutoff distance of the ML atoms, enabling them to learn semi-local electrostatic effects (35, 36). However, this strategy introduces a challenge: the dimension of features dramatically rises with the large number of surrounding MM atoms and makes it difficult to adequately sample the physical space. To reduce the dimensionality challenge, Yang and co-workers found that the electrostatic potential from the MM atoms is an effective collective variable for the neural network to learn the polarization of QM atoms (37, 38). With the same philosophy, recent works have used similar descriptors for MM environment and applied ML/MM models to a wide range of condensed phase properties prediction and energy calculation (39-42).

The second route involves calculating a correction term for polarization effects separate from the internal QM energy in vacuo. However, calculating this additional correction term remains challenging, as it involves theoretical modeling of the QM charges fluctuation due to the electrostatic environment, with the current studies limited to linear response (29, 43-45). Further comparison of these two routes will be discussed in section 4.1.

In this study, we developed a highly accurate ML/MM multiscale model with electrostatic embedding following the first route, utilizing its ability to effectively model charges perturbations beyond linear response while maintaining computational efficiency. We have followed our previous work (37) to train the neural network with external electrostatic potential being the descriptor for the MM environment. Taking the advanced architecture of the E(3)-equivariant NequIP model, we developed the **N**eural **e**quivariant **po**larizable **I**nteratomic **P**otentials (NepoIP) neural network and implemented the NepoIP/MM multiscale simulation in the OpenMM package (46). The performance of NepoIP/MM in MD simulation is validated through stable converged sampling of the periodic solvated alanine dipeptide system. We further demonstrated that our NepoIP model can be trained to be transferable across different MM environments, whether in water or in protein. The success of NepoIP/MM in these tasks forms the foundation for addressing the current limitations in simulations of large biomolecular systems.

## 2. Methods

### 2.1 Model Architecture

The architecture of NequIP was selected as our foundation due to its high data efficiency and accuracy. For an isolated molecule composed of *n* atoms, the original model embeds the input of atomic numbers $(Z_1, …, Z_n)$ into *n* feature vectors through a trainable self-interaction layer, which then go through the graph convolution layers in interaction blocks with edge information derived from the atom coordinates $(r_1, …, r_n)$. As in the case of ML/MM, electrostatic potential from the MM environment $(V_1, …, V_n)$ on each ML atom is incorporated to provide the ML model with environment information. For ML atom $i$, its electrostatic potential from the MM environment in non-periodic case is computed as:

$$V_i = \sum_{j \in MM} Q_j / |r_i - r_j|, \tag{1}$$

where $Q_j$ is the partial charge of MM atom $j$ given by its force field charge parameter, and $r_i$ and $r_j$ are coordinates of the ML atom $i$ and MM atom $j$. For the periodic case, the Ewald-summation of $V_i$ is formulated in Supporting Materials A1.

The external electrostatic potentials $(V_1, …, V_n)$ are embedded in the same way as $(Z_1, …, Z_n)$ and the features representing environment information are added to the internal features of the ML atoms. Since the model implicitly learns the partial charge fluctuation caused by the MM environment through the external electrostatic potentials, which represents a polarizable force field, it is termed as Neural equivariant polarizable Interatomic Potentials (NepoIP). The overview of NepoIP model is given in Fig. 1.

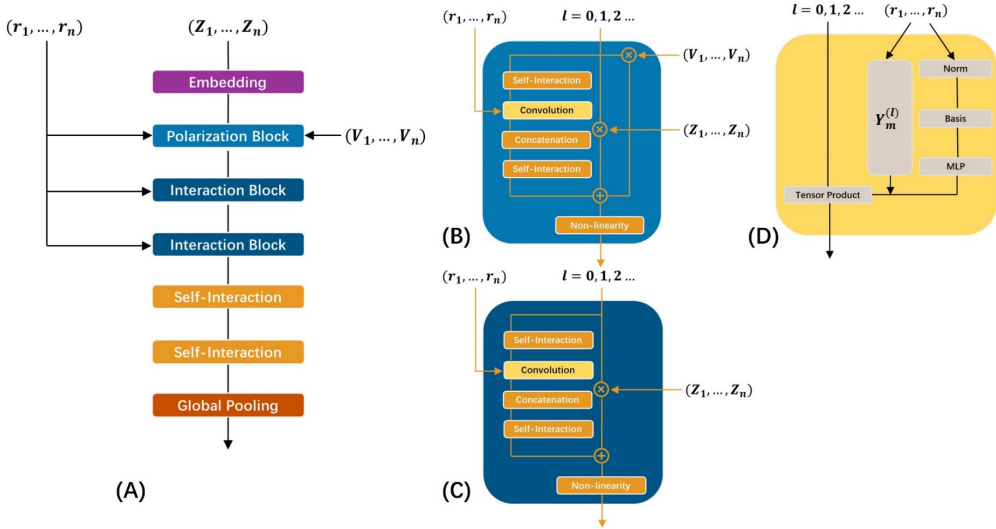

**Fig. 1**. Overview of the NepoIP model. (A) The embedded features are first refined through a polarization block, then refined through a series of interaction blocks. An output block then generates atomic energies, which are pooled to give the total predicted energy. (B) The polarization block, introducing a key difference from NequIP's

original interaction block by incorporating a tensor product between the neighbouring features and the external electrostatic potential. (C) The original interaction block, containing the convolution. (D) The convolution combines the product of the radial function and the spherical harmonic projection of the unit vector with neighbouring features via a tensor product.

We have also developed another model, termed as NepoIP[0], which incorporates the external electrostatic potential on ML atoms by embedding them in the same way as the atomic numbers (Fig. S1). NepoIP[0] is slightly computationally faster but also has slightly higher energy error than NepoIP, with their performance compared in Table S1. Both NepoIP[0] and NepoIP do not contain any new type of operation other than those in NequIP, the mathematical details of all the operations in the model have been well documented in the original literature (22). The E(3)-equivariance of the graph neural network is preserved as the electrostatic potential is used as a scalar attribute for each node on the graph.

## 2.2 Prediction of the NepoIP/MM Multiscale Model

In this section, we will further explain the predicted energy and forces of NepoIP in the context of ML/MM system, which is inherently different from the prediction for pure ML system.

MD simulation with machine learned potential is known to be unstable where the simulated systems may collapse on a short timescale (28). Although recent efforts have been made to reduce the unphysical samples by fine-tuning the NequIP model parameters through a reweighting scheme (23), the collapse of simulation only gets postponed rather than avoided. On the other hand, we found it is effective to guarantee the stability of simulation by using the model to predict the energy difference between QM and MM (Δ-machine learning) as in Charmm-NN (31), since the MM energy functions have extremely large penalties for the unphysical conformations. With the Δ-machine learning strategy, the total energy of a QM/MM system in our embedding scheme is thus decomposed as:

$$E_{total} = E_{QM}(QM) + E_{QM/MM}(QM/MM) + E_{MM}(MM)$$
$$= [E_{QM}(QM) + E_{QM/MM}(QM/MM) - E_{MM}(QM) - E_{MM}(QM/MM)]$$
$$+ E_{MM}(QM) + E_{MM}(QM/MM), \quad (2)$$

where $E_{QM}(QM)$ and $E_{QM/MM}(QM/MM)$ are the QM energy of the QM region and the QM/MM coupling energy between the QM and MM regions. $E_{MM}(QM)$ and $E_{MM}(MM)$ are the energy of the QM region and that of the MM region calculated at the MM level. $E_{MM}(QM/MM)$ is the coupling energy between the QM and MM regions calculated at the MM level.

In the electrostatic embedding scheme of QM/MM, the coupling energy $E_{QM/MM}(QM/MM)$ is calculated as:

$$E_{QM/MM}(QM/MM) = E^{elec}_{QM/MM}(QM/MM) + E^{vdw}_{QM/MM}(QM/MM)$$

$$= \sum_{i \in QM} \sum_{j \in MM} \frac{Z_i Q_j}{|r_i - r_j|} - \sum_{j \in MM} Q_j \int \frac{\rho(r')dr'}{|r' - r_j|} + E^{vdw}_{MM}(QM/MM), \quad (3)$$

where $r'$ represents electron coordinates, $\rho(r')$ is the electron density, $r_i$ and $Z_i$ represent the coordinate and neclear charge (atomic number) of QM atom $i$ respectively, $r_j$ and $Q_j$ represent the coordinate and partial charge of MM atom $j$ respectively. The Van der Waals interaction between the QM and MM regions remains to be calculated by the Lennard-Jones potential as the same as the MM level. We thus define the machine learning energy as the terms within the square bracket in Eq. (2); namely,

$$E_{ML} = E_{QM}(QM) + E_{QM/MM}(QM/MM) - E_{MM}(QM) - E_{MM}(QM/MM)$$
$$= E_{QM}(QM) + E_{QM/MM}^{elec}(QM/MM) - E_{MM}(QM) - E_{MM}^{elec}(QM/MM). \quad (4)$$

The QM/MM total energy in Eq. (2) can then be transformed into a simple ML/MM total energy as:

$$E_{total} = E_{MM}(MM) + E_{MM}(QM) + E_{MM}(QM/MM) + E_{ML}$$
$$= E_{MM} + E_{ML}, \quad (5)$$

where $E_{MM}$ represents the MM energy of the whole system. Eqs (4-5) thus define our $\Delta$ machine learning approach for ML/MM electrostatic embedding.

Under this potential, the ML force on an ML atom $i$ is:

$$\frac{\partial E_{ML}(r)}{\partial r_{ML,i}} = \frac{\partial E_{ML}(r_{ML}, V_{ML})}{\partial r_{ML,i}} + \sum_{j \in ML} \frac{\partial E_{ML}(r_{ML}, V_{ML})}{\partial V_{ML,j}} \frac{\partial V_{ML,j}}{\partial r_{ML,i}}, \quad (6)$$

where $r$ represents the coordinates of the whole system, $r_{ML}$ and $V_{ML}$ represent the coordinates and electrostatic potentials on the ML atoms respectively. The first term disregards the dependence of the electrostatic potential $V_{ML}$ to the coordinates $r_i$, and it exists only when $i$ is an ML atom.

The ML force on an MM atom $i$ is:

$$\frac{\partial E_{ML}(r)}{\partial r_{MM,i}} = \sum_{j \in ML} \frac{\partial E_{ML}(r_{ML}, V_{ML})}{\partial V_{ML,j}} \frac{\partial V_{ML,j}}{\partial r_{MM,i}}. \quad (7)$$

It should be noted that although there exists such a machine learned correction of forces on the MM region, the model is not directly trained to them.

### 2.3 Dataset Construction

The datasets constructed in this work are mainly for peptides in water as shown in Fig. 2.A, where the peptide molecule is defined as the QM region. We term them as the peptide-in-water datasets. To construct the datasets, MD simulation with Amber ff99SB force field was conducted to sample enough conformations. Then, QM/MM energy and forces were calculated for the samples and the reference energy expressed in Eq. (3) and its corresponding forces on the QM atoms were extracted.

On the other hand, in the realistic condition of biological systems, the water environment of an amino acid from a protein could vary from fully solvated when it appears on the surface area to not solvated at all when it is buried in the protein core. The external electrostatic potential on the amino acid in these two cases could be entirely different, thus we constructed the peptide-in-protein type of QM/MM datasets where the peptide molecule is defined as the QM region. The alanine dipeptide was first manually placed in the folding core of a cofactor-binding protein PS1 (PDBID: 5TGY) by replacing the original cofactor. The PS1 protein was *de novo* designed to bind its non-natural porphyrin cofactor with optimized stability (47) and provides an ideal folding core to place our peptide as shown in Fig. 2.B. MD sampling was conducted for the solvated peptide-in-protein system while all solvent and ions were removed when generating the reference dataset to mimic an idealized pure protein environment.

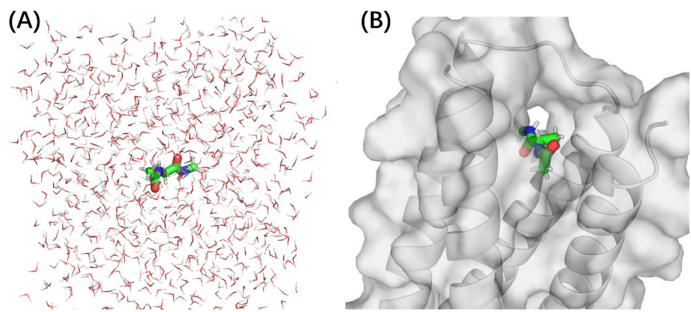

**Fig. 2.** Two types of QM/MM dataset: A. peptide-in-water B. peptide-in-protein.

For dataset construction, all MD sampling and QM/MM calculation were done in the Amber software. While the MD simulations were performed with periodic boundary conditions, the periodicity was removed when generating the reference dataset to prevent the model from potentially depending on any PBC settings (e.g. the cutoff scheme).

The QM/MM dataset for training NepoIP is inherently different from the QM gas phase datasets, since more information from the MM background is needed and the prediction has to be rigorously matched with its QM/MM counterparts as illustrated above. The workflow of extracting the reference data along with technical details of the MD sampling are described in Supporting Materials A2 & A3.

### 2.4 Multiscale MD Simulation

MD simulation at nanosecond timescale with both the NepoIP/MM and QM/MM models were done for the periodic solvated alanine dipeptide system. For all multiscale MD simulation, the initial system was taken from the last frame of the equilibration step under the ff99SB force field (48). For each model, we conducted 8 parallel trajectories of 2ns in the NPT ensemble with randomly generated intial velocity at temperature of 300 K. The periodic boundary condition was applied for both the QM (or ML) and the MM region with the direct space cutoff distance for Ewald summation set as 9 Å. The timestep is 2 fs and the bonds including H atoms are constrained for all simulation.

In our current implementation of NepoIP/MM, the simulation speed for periodic systems with Ewald summation applying for the ML atoms highly depends on the Ewald error tolerance $\delta$ (defined in Supporting Materials A1). The influence of setting different $\delta$ on simulation speed and prediction error is shown in Fig. S2. In consideration of the trade-off between simulation speed and error, we used $\delta = 0.005$ for the computation of electrostatic potential on ML atoms.

From the MD simulation trajectories, the distribution of the backbone dihedrals $\varphi$, $\psi$ were extracted to get the Ramachandra plot. In the analysis of the Ramachandra plot, the classification of secondary structures was defined following the usual tradition (49). We also calculated the ensemble averaged $^3J(H^N H^\alpha)$ couplings through the Karplus Equation with the 'solution' coefficients (50).

## 3. Results

### 3.1 Reproduction of the QM/MM Potential Energy

We begin by comparing the accuracy of NepoIP to non-polarizable models in reproducing the QM/MM energy of peptide-in-water. The reference QM theory was at the semi-empirical DFTB level with dispersion correction (51), with TIP3P water model (52) for the MM region. The reference energy includes the self-energy of the solute and its interaction with water molecules.

Three categories of models are taken into comparison. 1. ML/MM models with mechanical embedding. These models explicitly treat solute-solvent interaction at the MM level while ignoring the polarization effects. Instead of training any ML force field on the gas-phase dataset, we directly evaluated the QM/MM with mechanical embedding, representing the limit for this category. 2. ML non-polarizable force fields trained to the QM/MM energy including solute-solvent interaction. This draws inspiration from the strategy in work (28), where NequIP was trained against the pure MM energy of the solute and its interactions with the solvent, implicitly capturing solvation effects within a force field. In our case of QM/MM, the NequIP force field goes further by implicitly modeling both the solvation and polarization effects. 3. The polarizable force field NepoIP, which explicitly treats both the solute-solvent interaction and polarization effects.

The results shown in Fig. 3 indicate that the upper limit of ML/MM with mechanical embedding is even worse than implicitly learning all the solvation effects by NequIP. Meanwhile, NepoIP significantly reduced the errors of NequIP, highlighting the necessity and effectiveness of introducing the electrostatic potential to ML force fields for accurately reproducing the QM/MM energy. Besides, the performance of NepoIP is further increased when it is trained on larger datasets consisting of 50k, and 100k data points as shown in Table S2.

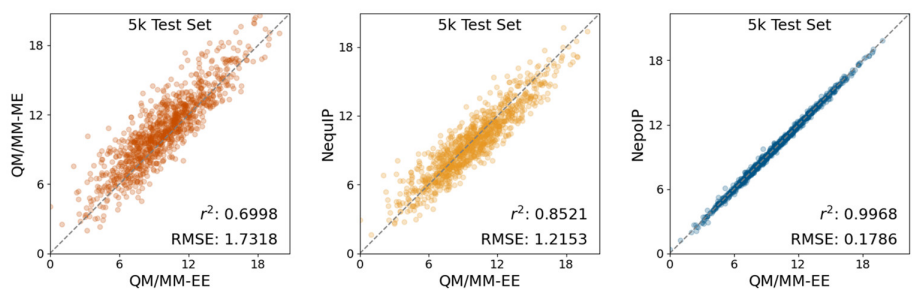

**Fig. 3.** Performance of different models in reproducing the QM/MM energy (in kcal/mol) of peptide-in-water. QM/MM-EE denotes QM/MM with electrostatic embedding and QM/MM-ME denotes mechanical embedding. Models (except QM/MM-ME) were trained on the peptide-in-water dataset consisting of 5k data points, with 1k data points reserved for testing. Correlation coeffients and Root-Mean-Squared-Errors are presented at the corner for each model.

### 3.2 Reproduction of the QM/MM MD Simulation

One of the most ardent dreams of biophysicists is to simulate a biomolecular system with the accuracy of QM efficiently enough to get quantitative insights (27). As a first step in pursuing this dream, molecular dynamics simulations with nanosecond scale were conducted for the alanine-in-water system to gain insights into its secondary structure propensity.

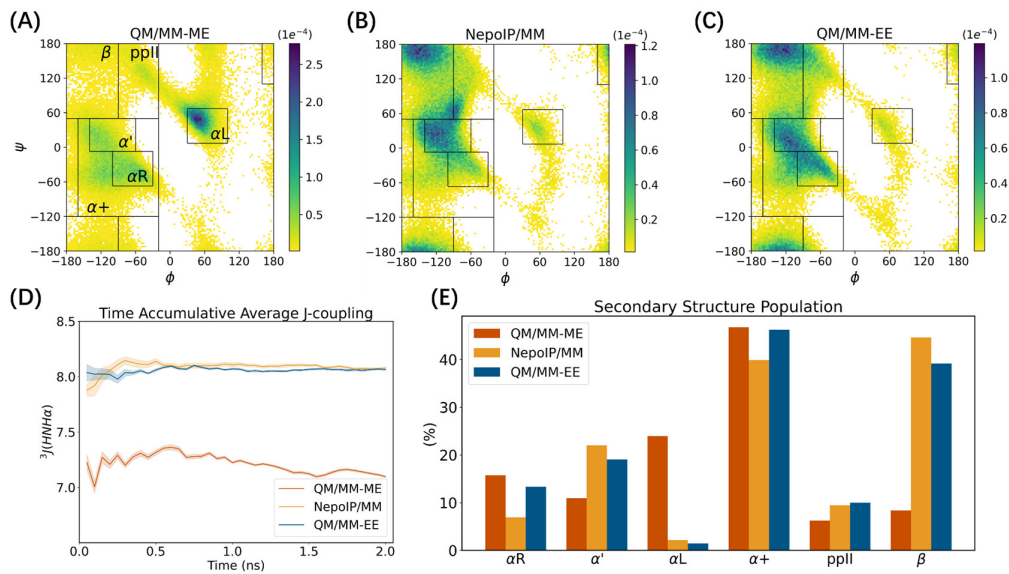

**Fig. 4.** Performance of NepoIP/MM in reproducing the QM/MM MD simulation. The 8*2ns simulations were performed under (A) QM/MM with mechanical embedding, (B) NepoIP/MM, and (C) QM/MM with electrostatic embedding. The secondary structure classes are labeled on (A) including $\beta$ sheet, ppII, extended right hand helix $\alpha+$ (containing $\alpha R$ and $\alpha'$), and left hand helix $\alpha L$. (D) The time accumulative average $^3J(H^N H^\alpha)$ value in Hz. The average values among 8 parallel trajectories are shown, with shaded regions indicating the standard deviation among the trajectories. (E) The average percentage population of secondary structures from the simulations.

As shown in Fig.4, the results of QM/MM with mechanical embedding are distinctly different from those of QM/MM with electrostatic embedding, indicating that machine learning force fields trained against the gas phase datasets are not readily to be used in solution phase simulation without modeling the polarization effects.

The NepoIP/MM simulations remained stable throughout the nanoscale time period, with no collapse occurred. It produces a Ramachandran plot closely aligned with the results of direct QM/MM simulations with electrostatic embedding and gives reasonable estimation of the secondary structure population. Furthermore, the converged $^3J(H^N H^\alpha)$ average value (8.078 Hz) matches well with the QM/MM-EE reference (8.065 Hz).

The simulation speed of NepoIP/MM is ~360 ps/day on 8 cores Intel Xeon 2.40GHz for this system with 22 ML atoms and ~3k MM atoms, which is acceptable but heavily restrained by the calculation of long range electrostatic potential with periodicity as shown in Fig S2. Better parallelization of the Ewald summation, or the implementation of more efficient methods, such as the particle meshed grid Ewald (53), and the random batch Ewald (54) recently proposed, are expected to significantly accelerate our simulation.

### 3.3 Transferability Across MM Force Fields

To validate that the same NepoIP/MM model can still be used to reproduce the QM/MM forces when the MM force field is changed, we built additional test datasets for two other water models with the samples from the peptide-in-water test set containing 1k data points. As shown in Table 1, the NepoIP model trained against the TIP3P water model can be used together with other water models with generally the same accuracy.

**Table 1.** Transferability of NepoIP to other MM water models

| Water model for Training | Water model for Testing | $Q_o$* (e) | RMSE | |
|---|---|---|---|---|
| | | | E Test (kcal · mol$^{-1}$) | F Test (kcal · mol$^{-1}$Å$^{-1}$) |
| TIP3P | TIP3P | -0.834 | 0.179 | 0.287 |
| | TIP3P-FB | -0.848 | 0.177 | 0.276 |
| | OPC3 | -0.895 | 0.192 | 0.282 |

*$Q_o$: The point charge of the oxygen atom in water

This direct transferability across MM force fields is ensured because, in our modeling approach, the MM background is treated as an electric field for the neural networks. As long as the distribution of the electrostatic potential generated by the MM background remains within the range covered by our training dataset, the model can accurately capture the corresponding polarization effects. Such a property is also an advantage of using electric field over atomwise descriptors for the MM background.

### 3.4 Transferability Between the Water and Protein Environment

It is known that the electrostatic environment in proteins differs significantly from that in pure water (55). This is also revealed in Fig. 5, where the external electrostatic potential (ESP) on the ML atoms exhibits distinctly different distributions between the water and protein environments.

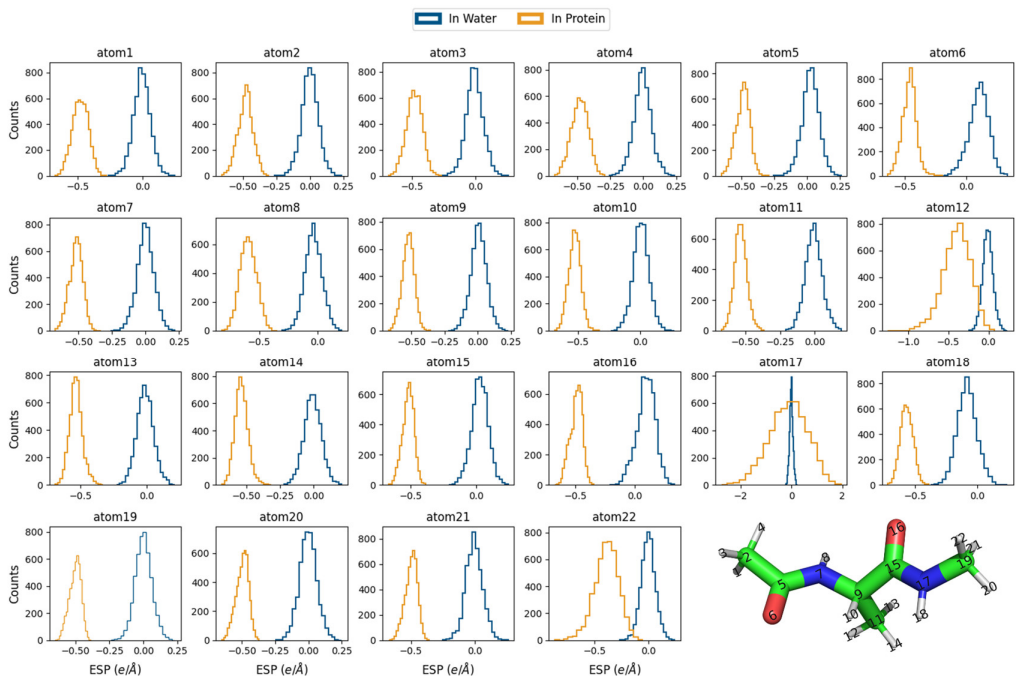

**Fig. 5.** Distribution of the external electrostatic potential on dipeptide atoms in 100 ns MD samplings. Snapshots were taken every 20 ps, yielding 5,000 data points for each environment. Atom indices are labeled on the dipeptide structure for reference.

The ML force field trained specifically to one environment is not directly applicable to the other (the large errors are not shown). This is consistent with our expectation since the range of ESP is totally different from the two environments. Here, we merged the two datasets to train a single model and test it on the original test sets. As shown in Table 2, although the test errors of this single model increase for the peptide-in-water system and decrease for the peptide-in-solvent system, they remain comparable to the original errors of models trained specifically for each environment. Therefore, by covering the distribution of ESP in the dataset, it is possible to train a general ML polarizable force field for totally different MM environments.

**Table 2.** Transferablility across totally different MM environments

| Training dataset | Test dataset | RMSE (kcal·mol$^{-1}$ or kcal·mol$^{-1}$Å$^{-1}$) | | | |
|---|---|---|---|---|---|
| | | E Train | F Train | E Test | F Test |
| ALA/Water-4k | ALA/Water-1k | 0.175 | 0.181 | 0.179 | 0.287 |
| ALA/Protein-4k | ALA/Protein-1k | 1.660 | 0.400 | 1.670 | 0.497 |
| ALA/Merged-8k | ALA/Water-1k | 0.975 | 0.303 | 0.255 | 0.337 |
| ALA/Merged-8k | ALA/Protein-1k | 0.975 | 0.303 | 1.377 | 0.407 |

## 4. Discussion

### 4.1 Learning Polarization and Deriving Polarization

NepoIP/MM demonstrates the ability of neural networks to learn the polarization effects induced by the MM point charges in a QM/MM model using electrostatic embedding. By capturing these polarization effects, NepoIP/MM accurately reproduces the conformational distributions observed in biomolecular simulations, positioning it as a computationally efficient alternative to traditional QM/MM approaches in this context.

Similar attempts of simulating peptides in solution by ML/MM model with electrostatic embedding were made in recent studies including the EMLE (43) and the ANI-MBIS (44) methods. In these studies, the polarization effects are derived analytically (the 'second route' introduced in section 1) rather than integrated into the neural network. The advantage of this strategy is that it allows the neural networks trained against the gas phase QM dataset of the small molecules to be directly used in an ML/MM system without further training.

The Ramachandra plot of the alanine dipeptide produced by our NepoIP/MM model and the full EMLE method (referred to as EMLE-total in the original literature) (43) both successfully captured the patterns of their corresponding QM/MM reference, while both of them still exhibit noticeable errors. On the other hand, in the full EMLE method, the reported error of the embedding model itself (excluding the error from the neural network for gas-phase energies) is already 1.83 kcal/mol for the alanine-in-water system. This error arises from the approximation by using the Thole model (56) to derive only the linear response of QM charges. In contrast, our NepoIP model achieves an overall energy RMSE of less than 0.1 kcal/mol on our test dataset of the same system, indicating that it successfully captures charges polarization induced by the MM environment beyond the linear response calculated from the Thole model.

In the case of the ANI-MBIS method, a simpler approximation was made to calculate the linear response through the multiplication of fixed atomic polarizability parameters with the external electric field (29, 44). Although the ML/MM free energy surface of alanine tetrapeptide showed roughly reasonable minima location, the reasonability of distribution was not assessed by comparing with the corresponding $\omega$B97x/MM (57) results.

Apart from simulation accuracy, analytically deriving the polarization effects during ML/MM simulation must be done on-the-fly, which inevitably adds to the computational cost on top of the neural network calculations. Although extra computation of electrostatic potential is also introduced in our current implementation of NepoIP/MM simulation, it comes at the cost of pure MM calculations, and can be incorporated into the computation of MM electrostatic interaction in principle to avoid the current repetitive computation.

Therefore, we suggest that learning the polarization may add difficulty to train the NN but could result in lower error and lower computational cost during simulations, whereas deriving the polarization presents the opposite scenario. Moreover, the advantages of learning polarization are expected to grow in the future, as the computational resources for generating larger dataset and training NN are rapidly growing, which could trivially improve the prediction accuracy. Besides, the computational efficiency of NN can benefit from the development of more efficient architectures and advancements in GPU technology. For example, the SO3krates neural network (58) has been recently developed to achieve accuracy comparable to NequIP but operates at a significantly faster speed. However, enhancing the accuracy of deriving polarization is a non-trivial challenge when the algorithm complexity is considered.

**4.2 Challenges for a General Model of Large Biomolecular Systems**

Although we have showed that NepoIP/MM can reveal the dynamics of small peptides at the QM/MM level, moving toward proteins remains a challenge. First, it will be too computationally demanding to generate datasets where an entire protein is treated at the QM level and sample its conformational space. Second, the molecular interactions can vary significantly in different protein systems, making it difficult to develop a general ML model that can be transferable among different proteins. For examples, recent attempts have been made to build a single transferable ML force field trained on extensive datasets that cover a diverse range of chemical systems (59, 60). However, noticeable errors in the Ramachandra plot of simulation based on the ML model exist even for small peptides in *vacuo*, comparing with the QM reference.

To address these challenges, Yang and coworkers have developed the residue-based systematic molecular fragmentation (rSMF) method (61) and later built a general ML protein force field, Charmm-NN, based on rSMF (31). The method partitions a protein into residue-level fragments and reconstructs the total energy from these fragments. This method enables the construction of a protein force field by training only a limited number of neural networks, each corresponding to a specific fragment type. As a result, the force field is inherently generalizable across different proteins. The method so far only supports mechanical embedding. This combination of fragmentation with ML in the ML/MM embedding as developed in (31, 61) allowed for the first time the construction of general force fields for biomolecules.

The recent AI$^2$BMD method (62) employs the same rSMF fragmentation approach and ML/MM mechanical embedding to develop a general protein force field. It has an enhanced MM environment using the AMOEBA polarizable force field (63). However, the ML model is still trained on protein fragment energies in the gas phase, neglecting the polarization effects of the ML region induced by the MM environment.

As is well known, the electronic polarization in the QM subsystem due to the MM environment is commonly used and plays a key role in the accuracy of the QM/MM simulations (15, 16, 34, 64-66). The same should be expected of a ML/MM force field. To build a general ML force field with the inclusion of polarization effects of the ML subsystem, a fragmentation method that supports electrostatic embedding is expected in the future, for which the present work paves the way.

## 5. Conclusion

To achieve biomolecular dynamics simulation with neural networks potential at the QM/MM level with electrostatic embedding, we have developed the NepoIP model, learning polarization effects based on the external electrostatic potential on the QM atoms. The necessity of incorporating the polarization effects is revealed by the significant difference between QM/MM-ME and QM/MM-EE potential energy and MD distribution. NepoIP/MM accurately reproduced the conformation distribution of the alanine dipeptide in solution.

By using the electrostatic potential as a collective variable, we have reduced the dimension of the descriptor for MM environment to $n$, only the number of ML atoms. Learning from this physical quantity, NepoIP is transferable across different MM force fields and MM environments when the distribution of external electrostatic potential on the ML atoms remains in the range covered by the training data. Presently, accurately modeling interactions in large protein systems at the quantum mechanics level by a general machine learning force field remains a challenging task. Developing a proper fragmentation scheme to model biomolecules in solution with machine learning polarizable force field is a future direction.

## Acknowledgement

We acknowledge support from the National Institute of Health (R01-GM061870).

## Data and Code Availability

The NepoIP/MM model is implemented in OpenMM through the OpenMM-Torch API. The source codes of NepoIP/MM, scripts for NepoIP/MM simulation and datasets used to train NepoIP are available on https://github.com/Yang-Laboratory/NepoIP

# Supporting information

## A1. Ewald-summation of external electrostatic potential in periodic ML/MM systems.

For periodic ML/MM systems, the external electrostatic potential on each ML atom from the MM environment is calculated by subtracting the contribution of the ML atoms in the central box from the electrostatic potential of the entire periodic system. The external electrostatic potential on ML atom $i$ is:

$$V_i = V_i^{ewald} - \sum_{j \in ML}{}' \frac{Q_j}{r_{ij}},$$

where $V_i^{ewald}$ is the electrostatic potential on atom $i$ from the entire periodic system, $Q_j$ is the MM point charge of atom $j$, $r_{ij}$ is the distance between atom $i$ and $j$, and the $'$ symbol denotes the exclusion of the term $i = j$.

The $V_i^{ewald}$ is calculated as the sum of the direct space term, the reciprocal space term, and the self-energy term:

$$V_i^{ewald} = V_i^{dir} + V_i^{rec} + V_i^{self}.$$

1. The direct space term:

$$V_i^{dir} = \sum_{j \in \{\mathcal{N}_i\}} Q_j \frac{erfc(\alpha r_{ij})}{r_{ij}},$$

where $\{\mathcal{N}_i\}$ is the neighbor atom list of atom $i$ within the cutoff distance, $r_{cutoff}$, which is user-specified in OpenMM. $erfc$ is the complementary error function, and $\alpha$ is a coefficient calculated from an user-specified error tolerance $\delta$ in OpenMM:

$$\alpha = \sqrt{-\log(2\delta)}/r_{cutoff}.$$

2. The reciprocal term:

$$V_i^{rec} = \frac{1}{\pi V} \sum_j Q_j \sum_{k \neq 0} \frac{\exp(-(\pi \mathbf{k}/\alpha)^2) \cos(2\pi \mathbf{k} \cdot (r_i - r_j))}{\mathbf{k}^2},$$

where $V$ is the volume of the periodic box. $\mathbf{k}$ is the wave vector:

$$\mathbf{k} = (k_1, k_2, k_3)/\mathbf{L},$$

where $k_1, k_2, k_3$ are all integers from $-k_{max}$ to $k_{max}$ except $(0, 0, 0)$ and $\mathbf{L}$ is the periodic box length vector. The value of $k_{max}$ for each dimension is selected to be the smallest value which gives an estimated error less than the user-specified error tolerance $\delta$. The error is estimated as:

$$\varepsilon = \frac{k_{max}\sqrt{L\alpha}}{20} \exp\left(-(\pi k_{max}/L\alpha)^2\right),$$

where $L$ is the periodic box length for a specific dimension. If the box is not cubic, $k_{max}$ can have different values along different dimensions.

3. The self-energy term:

$$V_i^{self} = -\frac{2\alpha}{\sqrt{\pi}} Q_i^2.$$

This computation of external electrostatic potential in periodic ML/MM systems is implemented in the OpenMM-Torch API using PyTorch as the programming framework. The intensive computation of the reciprocal term is parallelized through computing the energy from different $\mathbf{k}$ vectors in batch.

# Supporting information

## A2. Sampling of the QM/MM dataset

For both the peptide-in-water and peptide-in-protein datasets, we have performed 100 ns regular MD sampling under the Amber ff99SB force field and TIP3P water model with the following procedures.

For peptide-in-water, the alanine dipeptide was solvated in a water box with a 12 Å edge distance. For peptide-in-protein, the system was solvated in a water box with a 10 Å edge distance. The systems were minimized, heated up to 300 K at a constant volume, then equilibrated at a constant pressure of 1 bar before the final 100 ns production MD in the NPT ensemble. The direct space cutoff distance for PME is 10 Å throughout the procedures for both of the cases.

To enhance the performance of NepoIP/MM MD simulation, an umbrella sampling dataset was constructed for the peptide-in-water system. With settings being the same as regular MD, the umbrella sampling was conducted with the φ, ψ dihedrals dihedrals divided into windows of $30° \times 30°$, yielding $12 \times 12$ windows. For each window, after minimization, heating up to 300 K and equilibration at 1 bar, 20ns production simulation was conducted under harmonic constraint with force constant of $20 \text{ kcal} \cdot \text{mol}^{-1}\text{rad}^{-1}$ applying only outside of the window boundaries.

## A3. Extraction of the QM/MM reference data

QM/MM reference energy and forces were calculated for the sampled snapshots without periodicity nor any restraints (e.g. the SHAKE algorithm used in Amber as default) by rerunning the trajectories (imin=5 in Amber). Since we took the Δ-machine learning strategy to predict the correction of energy and forces by neural network, pure MM energy and forces were also calculated in the same way and were subtracted from those of QM/MM to get the final reference data. The external electrostatic potential on an ML atom was calculated according to eq. (1). The gradient $\frac{\partial V_j}{\partial r_i}$, needed for forces calculation, was taken from the analytical differentiation of eq. (1). This gradient is the electric field on an atom when $i = j$.

As a result, the final dataset comprises six components: (A) atom coordinates, (B) atomic numbers, (C) electrostatic potential on the atoms, (D) gradients of the electrostatic potential, (E) reference energy, and (F) reference forces. These components pertain exclusively to the ML atoms, dramatically reducing the dimension of input by excluding any MM atoms.

# Supporting information

**B. Supporting Figures**

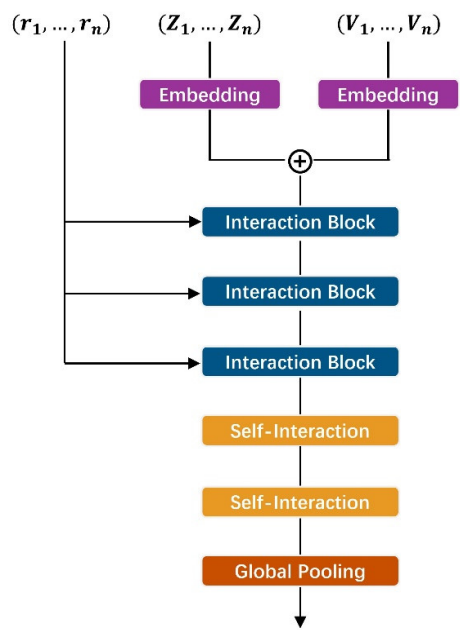

**Fig. S1.** Overview of the NepoIP$^0$ model. The atomic features embedded from atomic numbers and external electrostatic potential are merged and then refined through the same series of interaction blocks of NequIP. An output block then generates atomic energies, which are pooled to give the total predicted energy.

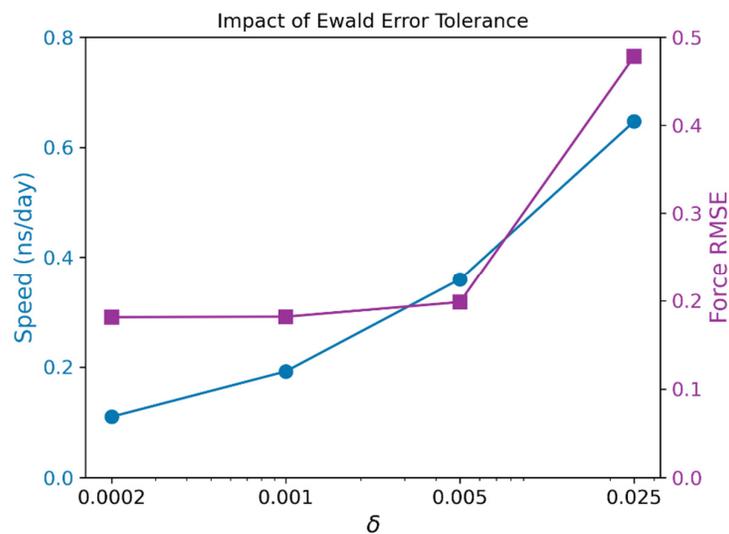

**Fig. S2.** Setting the Ewald error tolerance $\delta$ according to the trade-off between simulation speed and prediction error (kcal·mol$^{-1}$Å$^{-1}$). The speed increases steadily with higher tolerance. Force rmse is roughly converged with $\delta$ smaller than 0.005. Simulation speed is tested on 8 cores Intel Xeon 2.40GHz for periodic ala-in-water system. Force rmse is evaluated on 800 snapshots taken from 8*2ns NepoIP/MM simulation with the model trained on the 144k umbrella sampling dataset.

# Supporting information

## C. Supporting Tables

Table S1. Performance comparison of NepoIP$^0$ and NepoIP on ala-in-water 50k dataset

| Train/Test Split | Model | MD Speed* (ps/day) | RMSE (kcal·mol$^{-1}$ or kcal·mol$^{-1}$Å$^{-1}$) | | | |
|---|---|---|---|---|---|---|
| | | | E Train | F Train | E Test | F Test |
| 45k/5k | NepoIP$^0$ | ~366 | 0.0830 | 0.122 | 0.0840 | 0.131 |
| | NepoIP | ~360 | 0.0730 | 0.130 | 0.0741 | 0.140 |

*Tested on 8 cores Intel Xeon 2.40GHz for periodic ala-in-water system with $r_{cutoff} = 9$Å and $\delta = 0.005$

Table S2. Performance of NepoIP on ala-in-water datasets with increasing size.

| Dataset* | Train/Test Split | RMSE (kcal·mol$^{-1}$ or kcal·mol$^{-1}$Å$^{-1}$) | | | |
|---|---|---|---|---|---|
| | | E Train | F Train | E Test | F Test |
| 5k | 4k/1k | 0.175 | 0.181 | 0.179 | 0.287 |
| 50k | 45k/5k | 0.0730 | 0.130 | 0.0741 | 0.140 |
| 100k | 90k/10k | 0.0596 | 0.113 | 0.0595 | 0.118 |

* Samples in 5k, 50k, and 100k datasets are from the same 100ns MD sampling.